\documentclass[conference,hidelinks]{IEEEtran}
\IEEEoverridecommandlockouts
\usepackage{cite}
\usepackage{amsmath,amssymb,amsfonts}
\usepackage{algorithmic}
\usepackage{graphicx}
\usepackage{textcomp}
\usepackage{xcolor}
\usepackage[a4paper, total={184mm,239mm}]{geometry}
\def\BibTeX{{\rm B\kern-.05em{\sc i\kern-.025em b}\kern-.08em
    T\kern-.1667em\lower.7ex\hbox{E}\kern-.125emX}}
\usepackage{glossaries}
\usepackage{orcidlink}
\usepackage[outputdir=build,frozencache=true]{minted}
\usepackage[position=b]{subcaption}
\usepackage{cleveref}
\usepackage{tikz}
\usetikzlibrary{tikzmark,calc,positioning}

\newacronym[plural=WANs, firstplural={Wide Area Networks (WANs)}]{wan}{WAN}{Wide Area Network}
\newacronym[plural=WSNs, firstplural={Wireless Sensor Networks (WSNs)}]{wsn}{WSN}{Wireless Sensor Network}
\newacronym{simd}{SIMD}{Single Instruction Multiple Data}
\newacronym{os}{OS}{Operating System}
\newacronym{ble}{BLE}{Bluetooth Low-Energy}
\newacronym{wifi}{Wi-FI}{Wireless Fidelity}
\newacronym[plural=DVS, firstplural={Dynamic Vision Sensors (DVS)}]{dvs}{DVS}{Dynamic Vision Sensor}
\newacronym{ptz}{PTZ}{Pan-Tilt Unit}

\newacronym[plural=FLLs,firstplural=Frequency Locked Loops (FLLs)]{fll}{FLL}{Frequency Locked Loop}
\newacronym{dram}{DRAM}{Dynamic Random Access Memory}
\newacronym{fpu}{FPU}{Floating Point Unit}
\newacronym{fpss}{FPSS}{Floating Point Subsystem}
\newacronym{frep}{FREP}{Floating Point Repetition}
\newacronym{dma}{DMA}{Direct Memory Access}
\newacronym{ssr}{SSR}{Stream Semantic Register}
\newacronym{issr}{ISSR}{Indirection Stream Semantic Register}
\newacronym[plural=LUTs, firstplural={Lookup Tables (LUTs)}]{lut}{LUT}{Lookup Table}
\newacronym[plural=FPGAs, firstplural={Field Programmable Gate Arrays (FPGAs)}]{fpga}{FPGA}{Field Programmable Gate Array}
\newacronym{dsp}{DSP}{Digital Signal Processing}
\newacronym{mcu}{MCU}{Microcontroller Unit}
\newacronym{spi}{SPI}{Serial Peripheral Interface}
\newacronym{cpi}{CPI}{Camera Parallel Interface}
\newacronym{rf}{RF}{Register File}
\newacronym{fifo}{FIFO}{First-In First-Out Queue}
\newacronym{uart}{UART}{Universal Asynchronous Receiver-Transmitter}
\newacronym{raw}{RAW}{Read After Write}
\newacronym{waw}{WAW}{Write After Write}
\newacronym[plural=ISAs, firstplural={Instruction Set Architectures (ISAs)}]{isa}{ISA}{Instruction Set Architecture}
\newacronym{xbar}{XBAR}{crossbar}
\newacronym[firstplural=Scratch-Pad Memories (SPMs)]{spm}{SPM}{Scratch-Pad Memory}
\newacronym{ppa}{PPA}{Power Performance Area}
\newacronym{ipi}{IPI}{Inter-Processor Interrupt}
\newacronym[firstplural=Software-Generated Interrupts (SGIs)]{sgi}{SGI}{Software-Generated Interrupt}
\newacronym{pe}{PE}{Processing Element}
\newacronym{tcdm}{TCDM}{Tightly-Coupled Data Memory}
\newacronym{lsu}{LSU}{Load-Store Unit}
\newacronym{icache}{I\$}{Instruction Cache}
\newacronym{dcache}{D\$}{Data Cache}
\newacronym{wfi}{WFI}{Wait For Interrupt}
\newacronym{gpc}{GPC}{GPU Processing Cluster}
\newacronym{cpu}{CPU}{Central Processing Unit}
\newacronym{gpu}{GPU}{Graphics Processing Unit}
\newacronym{llc}{LLC}{Last-Level Cache}
\newacronym{sm}{SM}{Streaming Multiprocessor}
\newacronym[firstplural=Networks on Chip (NoCs)]{noc}{NoC}{Network on Chip}
\newacronym{csr}{CSR}{Control and Status Register}
\newacronym{ooo}{OoO}{Out-of-Order}

\newacronym{dfg}{DFG}{Data Flow Graph}
\newacronym{lcg}{LCG}{Linear Congruential Generator}
\newacronym{prn}{PRN}{Pseudo-Random Number}

\newacronym{ste}{STE}{Straight-Through-Estimator}

\newacronym[plural=PTUs, firstplural={Pan-Tilt Units}]{ptu}{PTU}{Pan-Tilt Unit}
\newacronym{mdf}{MDF}{Medium-density fibreboard}
\newacronym{cvat}{CVAT}{Computer Vision Annotation Tool}
\newacronym{coco}{COCO}{Common Objects in Context}
\newacronym{soa}{SoA}{State of the Art}
\newacronym{sf}{SF}{Sensor Fusion}

\newacronym{dl}{DL}{Deep Learning}
\newacronym{bn}{BN}{Batch Normalization}
\newacronym{FGSM}{FBK}{Fast Gradient Sign Method}
\newacronym{lr}{LR}{Learning Rate}
\newacronym{sgd}{SGD}{Stochastic Gradient Descent}
\newacronym{gd}{GD}{Gradient Descent}
\newacronym{llm}{LLM}{Large Language Model}

\newacronym{sta}{STA}{Static Timing Analysis}

\newacronym[plural=GPIOs, firstplural={General Purpose Inupt Outputs (GPIOs)}]{gpio}{GPIO}{General Purpose Input Output}
\newacronym[plural=LDOs, firstplural={Low Dropout Regulators (LDOs)}]{ldo}{LDO}{Low Dropout Regulator}

\newacronym{inq}{INQ}{Incremental Network Quantization}

\newacronym{CV}{CV}{Computer Vision}
\newacronym{EoT}{EoT}{Expectation over Transformation}
\newacronym{RPN}{RPN}{Region Proposal Network}
\newacronym{TV}{TV}{Total Variation}
\newacronym{NPS}{NPS}{Non-Printability Score}
\newacronym{STN}{STN}{Spatial Transformer Network}
\newacronym{MTCNN}{MTCNN}{Multi-Task Convolutional Neural Network}
\newacronym{YOLO}{YOLO}{You Only Look Once}
\newacronym{SSD}{SSD}{Single Shot Detector}
\newacronym{SOTA}{SOTA}{State of the Art}
\newacronym{NMS}{NMS}{Non-Maximum Suppression}
\newacronym{ic}{IC}{Integrated Circuit}
\newacronym{tcxo}{TCXO}{Temperature Controlled Crystal Oscillator}
\newacronym{jtag}{JTAG}{Joint Test Action Group industry standard}
\newacronym{swd}{SWD}{Serial Wire Debug}
\newacronym{sdio}{SDIO}{Serial Data Input Output}
\newacronym[plural=PCBs, firstplural={Printed Circuit Boards (PCB)}]{pcb}{PCB}{Printed Circuit Board}
\newacronym[plural=ASICs, firstplural={Application Specific Integrated Circuits}]{asic}{ASIC}{Application Specific Integrated Circuit}

\newacronym[plural=BNNs, firstplural={Binary Neural Networks (BNNs)}]{bnn}{BNN}{Binary Neural Network}
\newacronym[plural=NNs, firstplural={Neural Networks}]{nn}{NN}{Neural Network (NNs)}
\newacronym[plural=SCMs, firstplural={Standard Cell Memories (SCMs)}]{scm}{SCM}{Standard Cell Memory}
\newacronym{ann}{ANN}{Artificial Neural Networks}
\newacronym{ml}{ML}{Machine Learning}
\newacronym{ai}{AI}{Artificial Intelligence}
\newacronym{iot}{IoT}{Internet of Things}
\newacronym{fft}{FFT}{Fast Fourier Transform}
\newacronym[plural=OCUs, firstplural={Output Channel Compute Units (OCUs)}]{ocu}{OCU}{Output Channel Compute Unit}
\newacronym{alu}{ALU}{Arithmetic Logic Unit}
\newacronym{mac}{MAC}{Multiply-Accumulate}
\newacronym[firstplural={systems-on-chip (SoCs)}]{soc}{SoC}{system-on-chip}
\newacronym[firstplural={multi-processor systems-on-chip (MPSoCs)}]{mpsoc}{MPSoC}{multi-processor system-on-chip}

\newacronym{PGD}{PGD}{Projected Gradient Descend}
\newacronym{CW}{CW}{Carlini-Wagner}
\newacronym{OD}{OD}{Object Detection}

\newacronym{rrf}{RRF}{RADAR Repetition Frequency}
\newacronym{nlp}{NLP}{Natural Language Processing}
\newacronym{qam}{QAM}{Quadrature Amplitude Modulation}
\newacronym{rri}{RRI}{RADAR Repetition Interval}
\newacronym{radar}{RADAR}{Radio Detection and Ranging}
\newacronym{loocv}{LOOCV}{Leave-one-out cross validation}

\newacronym{bsp}{BSP}{Board Support Package}
\newacronym{ttn}{TTN}{The Things Network}
\newacronym{wip}{WIP}{Work in Progress}
\newacronym{json}{JSON}{JavaScript Object Notation}
\newacronym{qat}{QAT}{Quantization-Aware Training}

\newacronym{cls}{CLS}{Classification Error}
\newacronym{loc}{LOC}{Localization Error}
\newacronym{bkgd}{BKGD}{Background Error}
\newacronym{roc}{ROC}{Receiver Operating Characteristic}
\newacronym{frr}{FRR}{False Rejection Rate}
\newacronym{eer}{EER}{Equal Error Rate}
\newacronym{snr}{SNR}{Signal-to-Noise Ratio}
\newacronym{flop}{FLOP}{Floating-Point Operation}
\newacronym{fp}{FP}{Floating-Point}
\newacronym{fps}{FPS}{Frames Per Second}
\newacronym{oi}{OI}{Operational Intensity}
\newacronym{ipc}{IPC}{Instructions per Cycle}

\newacronym{gsc}{GSC}{Google Speech Commands}
\newacronym{mswc}{MSWC}{Multilingual Spoken Words Corpus}
\newacronym{demand}{DEMAND}{Diverse Environments Multichannel Acoustic Noise Database}

\newacronym[plural=SNNs, firstplural={Spiking Neural Networks (SNNs)}]{snn}{SNN}{Spiking Neural Network}
\newacronym[plural=DNNs, firstplural={Deep Neural Networks (DNNs)}]{dnn}{DNN}{Deep Neural Network}
\newacronym[plural=TCNs,firstplural=Temporal Convolutional Networks]{tcn}{TCN}{Temporal Convolutional Network}
\newacronym[plural=CNNs,firstplural=Convolutional Neural Networks (CNNs)]{cnn}{CNN}{Convolutional Neural Network}
\newacronym[plural=TNNs,firstplural=Ternarized Neural Networks]{tnn}{TNN}{Ternarized Neural Network}
\newacronym{ds-cnn}{DS-CNN}{Depthwise Separable Convolutional Neural Network}
\newacronym{rnn}{RNN}{Recurrent Neural Network}
\newacronym{gcn}{GCN}{Graph Convolutional Network}
\newacronym{mhsa}{MHSA}{Multi-Head Self Attention}
\newacronym{crnn}{CRNN}{Convolutional Recurrent Neural Network}
\newacronym{clca}{CLCA}{Convolutional Linear Cross-Attention}
\newacronym{bf}{BF}{Beamforming}
\newacronym{anc}{ANC}{Active Noise Cancellation}
\newacronym{agc}{AGC}{Automatic Gain Control}
\newacronym{se}{SE}{Speech Enhancement}
\newacronym{mct}{MCT}{Multi-Condition Training}
\newacronym{mcta}{MCTA}{Multi-Condition Training \& Adaptation}
\newacronym{pcen}{PCEN}{Per-Channel Energy Normalization}
\newacronym{mfcc}{MFCC}{Mel-Frequency Cepstral Coefficient}
\newacronym{asr}{ASR}{Automated Speech Recognition}
\newacronym{kws}{KWS}{Keyword Spotting}
\newacronym{odl}{ODL}{On-Device Learning}

\newacronym{nl-kws}{NL-KWS}{Noiseless Keyword Spotting}
\newacronym{na-kws}{NA-KWS}{Noise-Aware Keyword Spotting}
\newacronym{odda}{ODDA}{On-Device Domain Adaptation}
\newacronym{hpm}{HPM}{High-Performance Mode}
\newacronym{lpm}{LPM}{Low-Power Mode}

\newcommand{\ResultGeomeanEnergyImprovementOptI}{7}
\newcommand{\ResultGeomeanSpeedup}{4}
\newcommand{\ResultGeomeanSpeedupOverBaseMinus}{8}

\newcommand{\ResultGeomeanEnergyImprovementOptII}{10}
\newcommand{\ResultGeomeanEnergyImprovementOptIIOverBaseMinus}{9}

\setminted[]{
    escapeinside=||,
    style=arduino,
    breaklines,
    fontsize=\scriptsize
}

\definecolor{magma-yellow}{HTML}{fec22b}
\definecolor{magma-orange}{HTML}{ee7828}
\definecolor{magma-magenta}{HTML}{d1356d}

\Crefname{figure}{Fig.}{Figs.}

\begin{document}

% Shorten authors list in bibliography
\bstctlcite{IEEEexample:BSTcontrol}

\title{Late Breaking Results: A RISC-V ISA Extension\\for Chaining in Scalar Processors}

\ifdefined\blindreview
\else
\author{
    \IEEEauthorblockN{Luca Colagrande}
    \IEEEauthorblockA{
        \textit{Integrated Systems Laboratory (IIS)} \\
        \textit{ETH Zurich}\\
        Zurich, Switzerland \\
        colluca@iis.ee.ethz.ch\orcidlink{0000-0002-7986-1975}
    }
    \and
    \IEEEauthorblockN{Jayanth Jonnalagadda}
    \IEEEauthorblockA{
        \textit{D-ITET} \\
        \textit{ETH Zurich}\\
        Zurich, Switzerland \\
        jjonnalagadd@student.ethz.ch\orcidlink{0000-0001-7511-0579}
    }
    \and
    \IEEEauthorblockN{Luca Benini}
    \IEEEauthorblockA{
        \textit{Integrated Systems Laboratory (IIS)} \\
        \textit{ETH Zurich}\\
        Zurich, Switzerland \\
        lbenini@iis.ee.ethz.ch\orcidlink{0000-0001-8068-3806}
    }
}
\fi

\maketitle

\begin{abstract}
    Modern general-purpose accelerators integrate a large number of programmable area- and energy-efficient processing elements (PEs), to deliver high performance while meeting stringent power delivery and thermal dissipation constraints.
    In this context, PEs are often implemented by scalar in-order cores, which are highly sensitive to pipeline stalls.
    Traditional software techniques, such as loop unrolling, mitigate the issue at the cost of increased register pressure, limiting flexibility.
    We propose scalar chaining, a novel hardware-software solution, to address this issue without incurring the drawbacks of traditional software-only techniques.
    We demonstrate our solution on register-limited stencil codes, achieving \textgreater93\% FPU utilizations and a 4\% speedup and 10\% higher energy efficiency, on average, over highly-optimized baselines.
    Our implementation is fully open source and performance experiments are reproducible using free software.
\ifdefined\blindreview
    \footnote{https://hidden-for-double-blind-review.com}
\else
    \footnote{\url{https://github.com/colluca/snitch_cluster/tree/chaining}}
\fi
\end{abstract}


\vspace{-2pt}

\section{Introduction}

As power delivery and thermal dissipation challenge the development of ever more powerful high-performance computers \cite{villa2014}, designing energy-efficient architectures has become of paramount importance \cite{esmaeilzadeh2011}.
To this end, architectures featuring many energy-efficient in-order cores with shallow pipelines are favoured over multicores hosting a few power- and resource-hungry deeply-pipelined out-of-order processors \cite{huang2009}.

While in-order cores are more area- and energy-efficient, they are also more sensitive to pipeline stalls as a consequence of data hazards, leading to reduced performance.
% TODO: add reference to the above statement
Consider the RISC-V code in \Cref{fig:base}, implementing the vector operation $\mathbf{a} = b * (\mathbf{c} + \mathbf{d})$ on the state-of-the-art scalar in-order Snitch processor \cite{zaruba2021}.
To improve FPU utilization, we take advantage of the \gls{ssr} \gls{isa} extension of Snitch: registers \texttt{ft0} and \texttt{ft1} implicitly read the next value from the $\mathbf{c}$ and $\mathbf{d}$ arrays, respectively, and \texttt{ft2} implicitly writes to the next element in the $\mathbf{a}$ array.
In an in-order pipelined processor, the \gls{raw} dependency between the \texttt{fadd} and \texttt{fmul} instructions causes the latter to stall until the result from the prior is available, wasting useful cycles, three in the case of Snitch, equal to the number of pipeline stages in its \gls{fpu}.

\begin{figure}[t]
    % 1st column
    \begin{minipage}[t]{0.498\columnwidth}
        \vspace{0pt}
        \centering
        % 1st minipage
        \begin{minipage}[t]{0.8\textwidth}
            \begin{minted}{asm}
fadd.d ft3, ft0, ft1
fmul.d ft2, ft3, %[b]
addi %[i], %[i], 1
bneq %[i], %[len], -12
            \end{minted}
            \vspace{-6pt}
            \subcaption{}
            \vspace{12pt}
            \label{fig:base}
        \end{minipage}
        % 2nd minipage
        \begin{minipage}[t]{0.8\textwidth}
            \begin{minted}{asm}
fadd.d ft3, ft0, ft1
fadd.d ft4, ft0, ft1
fadd.d ft5, ft0, ft1
fadd.d ft6, ft0, ft1
fmul.d ft2, ft3, %[b]
fmul.d ft2, ft4, %[b]
fmul.d ft2, ft5, %[b]
fmul.d ft2, ft6, %[b]
addi %[i], %[i], 4
bneq %[i], %[len], -36
            \end{minted}
            \vspace{-6pt}
            \subcaption{}
            \vspace{6pt}
            \label{fig:unroll}
        \end{minipage}
    \end{minipage}%
    % 2nd column
    \vline
    \hfill
    \begin{minipage}[t]{0.45\columnwidth}
        \vspace{0pt}
        \centering
        \begin{minipage}[t]{\textwidth}
            \begin{minted}[linenos,xleftmargin=18pt,numbersep=6pt]{asm}
li   %[mask], 8
csrs 0x7C3, %[mask]
fadd.d ft3, ft0, ft1
fadd.d ft3, ft0, ft1
|\tikzmark{sinst0}|fadd.d ft3, ft0, ft1|\tikzmark{einst0}|
fadd.d ft3, ft0, ft1
fmul.d ft2, ft3, %[b]
|\tikzmark{sinst1}|                     |\tikzmark{einst1}|
fmul.d ft2, ft3, %[b]
fmul.d ft2, ft3, %[b]
fmul.d ft2, ft3, %[b]
|\tikzmark{sinst2}|addi %[i], %[i], 4|\tikzmark{einst2}|
bneq %[i], %[len], -36
csrs 0x7C3, x0
            \end{minted}
            \vspace{-4pt}
            \subcaption{}
            \label{fig:chaining}
            % TIKZ annotations
            \begin{tikzpicture}[
                remember picture,
                overlay
                ]

                \draw[magma-yellow,thick]
                ($(pic cs:sinst0)+(-0.15em,0.65em)$) -- % Top left corner
                ($(pic cs:einst0)+(0.15em,0.65em)$) -- % Top right corner
                ($(pic cs:einst0)+(0.15em,-0.25em)$) --  % Bottom right corner
                ($(pic cs:sinst0)+(-0.15em,-0.25em)$) -- % Bottom left corner
                cycle;

                \draw[magma-orange,thick]
                ($(pic cs:sinst1)+(-0.15em,0.65em)$) -- % Top left corner
                ($(pic cs:einst1)+(0.15em,0.65em)$) -- % Top right corner
                ($(pic cs:einst1)+(0.15em,-0.25em)$) --  % Bottom right corner
                ($(pic cs:sinst1)+(-0.15em,-0.25em)$) -- % Bottom left corner
                cycle;

                \draw[magma-magenta,thick]
                ($(pic cs:sinst2)+(-0.15em,0.65em)$) -- % Top left corner
                ($(pic cs:einst2)+(0.15em,0.65em)$) -- % Top right corner
                ($(pic cs:einst2)+(0.15em,-0.25em)$) --  % Bottom right corner
                ($(pic cs:sinst2)+(-0.15em,-0.25em)$) -- % Bottom left corner
                cycle;
            \end{tikzpicture}
        \end{minipage}
    \end{minipage}
    \caption{(a) Baseline and (b) unrolling-optimized code variants. (c) A sample execution trace of the chaining-optimized code. The colored issue slots are marked for reference in \Cref{fig:diagram}.}
\end{figure}

Loop unrolling and software pipelining techniques can be adopted to eliminate wasted cycles \cite{cardoso2017}, as illustrated in \Cref{fig:unroll}.
The downside of this approach is that it increases register pressure, limiting the effectiveness of this technique to small codes, to avoid register spills \cite{makino2021}.
Vector and out-of-order processors with renaming support mitigate this issue by implementing large physical \glspl{rf}, at the cost of much increased area and energy consumption.

In this work, we propose an alternative solution that enables the advantages of loop unrolling without increasing register pressure.
By utilizing the pipeline registers within the processor's functional units to store intermediate results, our approach is well-suited for integration into highly area- and energy-efficient cores.
While we demonstrate our implementation on the Snitch processor with its \gls{isa} extensions, the solution is not restricted to this particular design.

\section{Implementation}

\begin{figure}
    \centering
    \includegraphics[width=\columnwidth]{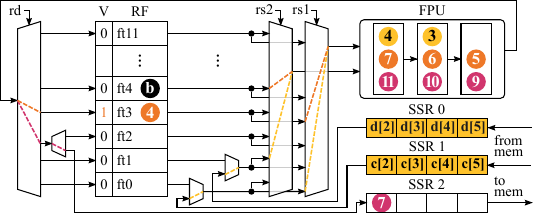}
    \caption{Block diagram of Snitch's \gls{fp} subsystem, illustrating the dataflow associated to the trace in \Cref{fig:chaining}. Elements in different colors represent snapshots at different moments in time, particularly at the respectively-colored issue slots in \Cref{fig:chaining}. Numbered tokens represent the outputs of the instructions at the respectively-numbered issue slots in \Cref{fig:chaining}.}
    \label{fig:diagram}
\end{figure}

The code in \Cref{fig:unroll} uses three additional architectural registers (\texttt{ft4}-\texttt{ft6}) to store the intermediate results of the \texttt{fadd} instructions, until they are consumed by the respective \texttt{fmul} instructions.
Our first insight is that the \gls{fpu} already provides enough storage for these intermediate results in its three pipeline stages, which led to unrolling the code by four in the first place.
By properly balancing the production and consumption rate of intermediate results by the \texttt{fadd} and \texttt{fmul} instructions, we can avoid allocating additional physical registers.
This dataflow, which conceptually involves \textit{chaining} the \texttt{fadd} and \texttt{fmul} instructions, is illustrated in \Cref{fig:diagram}.

As pipeline registers are not directly addressable, we need to extend the \gls{isa} to implement the desired dataflow.
We integrate a custom \gls{csr} (at address \texttt{0x7c3}) hosting a 32-bit mask (one bit per architectural register) to dynamically enable and disable chaining on the selected architectural \gls{fp} registers.
Enabling chaining alters instruction semantics: writeback to a chaining-enabled register is no longer enforced to be carried out atomically with operation execution.
There is thus no notion of \gls{waw} dependency between successive \texttt{fadd} instructions in the code in \Cref{fig:chaining}.
Instead, chaining-enabled registers are assigned \textit{FIFO semantics}, i.e. writes and reads to the register respectively imply push and pop operations from a logical \gls{fifo}, which is physically implemented by concatenating the chaining-enabled architectural register (\texttt{ft3} in our example) and the functional unit's pipeline registers.
Conversely, the traditional unrolling approach in \Cref{fig:unroll} implements this logical \gls{fifo} in the \gls{rf}, at the cost of precious architectural registers.
Thus, chaining benefits are increased for functional units with deeper pipelines.

Finally, we add a valid bit per architectural register to implement the backpressure mechanism avoiding that elements in the logical \gls{fifo} are overwritten before they are consumed in the occurrence of a stall, as illustrated by the orange issue slot in \Cref{fig:chaining}.

\section{Results}

We implement a Snitch cluster \cite{zaruba2021} with one compute core in GlobalFoundries' 12LP+ FinFET technology using Fusion Compiler 2023.12, with a target clock frequency of 1\,GHz.
Our extensions introduce negligible overheads, \textless2\% cell area increase and maximum frequency degradation, comparable to synthesis process variability margins.

All experiments are conducted in cycle-accurate RTL simulations using QuestaSim 2023.4.
Switching activities are extracted from post-layout simulations, and used for power estimation in PrimeTime 2022.03, assuming typical operating conditions of 25\,°C and 0.8\,V supply voltage.

We demonstrate the benefits of chaining on a set of stencil codes implemented in \cite{scheffler2024}.
We observe that the \texttt{box3d1r} and \texttt{j3d27pt} stencils are register-limited, and that, by applying chaining, enough registers can be freed to fully store the stencil coefficients in the \gls{rf}.

% TODO: add citation to baseline in legend
\begin{figure}[t]
    \centering
    \includegraphics[width=0.98\columnwidth]{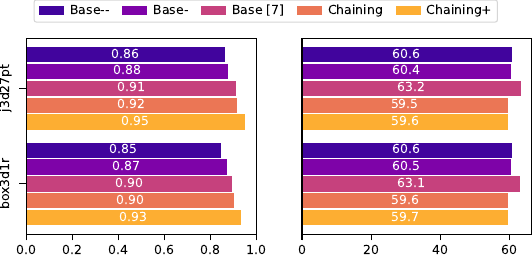}
    \vspace{-2pt}
    \caption{FPU utilization (left) and power consumption [mW] (right) for all code variants.}
    \label{fig:results}
\end{figure}

\Cref{fig:results} shows the results of our evaluation.
By moving the stencil coefficients to the \gls{rf}, we can get rid of repeated coefficient loads from memory.
In the implementation from \cite{scheffler2024}, these are mapped to an \gls{ssr}; thus, we do not directly benefit from eliminating issue cycles spent on explicit load instructions.
However, we observe a \ResultGeomeanEnergyImprovementOptI\% geomean improvement in energy efficiency, as repeated accesses to the L1 memory are avoided.
For comparison, the \texttt{Base--} implementation does not use an \gls{ssr} to load stencil coefficients.

Additionally, by freeing the \gls{ssr} previously employed in loading coefficients, chaining enables mapping writeback of stencil computation results to the freed \gls{ssr}, eliminating explicit store instructions.
We implement this optimization in \texttt{Chaining+} and \texttt{Base-}, which uses the \gls{ssr} freed in \texttt{Base--}.
Overall, chaining results in a \ResultGeomeanSpeedup\% geomean speedup and a \ResultGeomeanEnergyImprovementOptII\% geomean energy efficiency improvement over the highly-optimized baselines in \cite{scheffler2024}, and \ResultGeomeanSpeedupOverBaseMinus\% and \ResultGeomeanEnergyImprovementOptIIOverBaseMinus\% gains respectively over the direct comparison point \texttt{Base-}, reaching near-ideal \gls{fpu} utilizations above 93\%.

\section{Conclusion}

We presented a novel hardware and software solution to hide functional unit latencies in scalar in-order processors, without incurring increased register pressure, as with traditional software-only techniques.
We demonstrated our solution on a set of highly-optimized stencil codes, achieving a \ResultGeomeanSpeedup\% speedup and \ResultGeomeanEnergyImprovementOptII\% higher energy efficiency, on average, and \textgreater93\% FPU utilizations.
Our solution is lightweight and thus suited for integration into highly area- and energy-efficient cores.

\bibliography{paper}
\bibliographystyle{IEEEtran}

\end{document}